\documentclass[useAMS,usenatbib,usegraphicx,usehyperref]{mn2e}

\newcommand{\beq}{\begin{equation}}
\newcommand{\eeq}{\end{equation}}
\newcommand{\beqa}{\begin{eqnarray}}
\newcommand{\eeqa}{\end{eqnarray}}

\title[A generalised porosity formalism]{A generalised porosity formalism for isotropic and anisotropic effective opacity and its effects
on X-ray line attenuation in clumped O star winds}

\author[J.O. Sundqvist et al.]{Jon O. Sundqvist$^{1}$\thanks{E-mail: jon@bartol.udel.edu}, Stanley P. Owocki$^{1}$, 
David H. Cohen$^{2}$,\newauthor Maurice A. Leutenegger$^{3,4}$, and Richard H. D. Townsend$^{5}$\\
$^1$University of Delaware, Bartol Research Institute, Newark, Delaware 19716, USA\\
$^2$Swarthmore College, Department of Physics and Astronomy, Swarthmore, Pennsylvania 19081, USA\\
$^3$CRESST and X-ray Astrophysics Laboratory NASA/GSFC, Greenbelt, MD 20771, USA\\
$^4$Department of Physics, University of Maryland, Baltimore County, 1000 Hilltop Circle, Baltimore, MD 21250, USA\\
$^5$University of Wisconsin, Department of Astronomy, Madison, Wisconsin 53706, USA\\}

\begin{document}

\date{Accepted 2011-11-03. Received 2011-11-03; in original form 2011-10-02}

\pagerange{\pageref{firstpage}--\pageref{lastpage}} \pubyear{2002}

\maketitle

\label{firstpage}

\begin{abstract}
 
We present a generalised formalism for treating the
porosity-associated reduction in continuum opacity that occurs when
individual clumps in a stochastic medium become optically thick. As in
previous work, we concentrate on developing bridging laws between the
limits of optically thin and thick clumps. We consider geometries
resulting in either isotropic or anisotropic effective opacity, and,
in addition to an idealised model in which all clumps have the same
local overdensity and scale, we also treat an ensemble of clumps with
optical depths set by Markovian statistics. This formalism is then
applied to the specific case of bound-free absorption of X-rays in hot
star winds, a process not directly affected by clumping in the
optically thin limit. We find that the Markov model gives surprisingly
similar results to those found previously for the single clump model,
suggesting that porous opacity is not very sensitive to details of the
assumed clump distribution function.  Further, an anisotropic
effective opacity favours escape of X-rays emitted in the tangential
direction (the `venetian blind' effect), resulting in a 'bump' of
higher flux close to line centre as compared to profiles computed from
isotropic porosity models.  We demonstrate how this characteristic
line shape may be used to diagnose the clump geometry, and we confirm
previous results that for optically thick clumping to significantly
influence X-ray line profiles, very large porosity lengths, defined as
the mean free path between clumps, are required.  Moreover, we present
the first X-ray line profiles computed directly from line-driven
instability simulations using a 3-D patch method, and find that
porosity effects from such models also are very small. This further
supports the view that porosity has, at most, a marginal effect on
X-ray line diagnostics in O stars, and therefore that these
diagnostics do indeed provide a good `clumping insensitive' method for
deriving O star mass-loss rates.

\end{abstract}


\section{Introduction}
\label{intro}

Over the past years, it has become clear that in principle all
standard spectral mass-loss diagnostics of O stars are affected by
\textit{wind clumping}, i.e. by the small-scale wind inhomogeneities
that should arise naturally from a strong, intrinsic instability
associated with the radiative line-driving of these winds \citep[the
  line-driven instability, LDI, e.g.][]{Owocki88}. If neglected, such
wind clumping causes standard diagnostics such as H$_\alpha$ and
IR/radio free-free emission, which have opacities that depend on the
local wind density squared, to overestimate mass-loss rates \citep[for
  summaries, see][]{Puls08, Hamann08}.\footnote{Diagnostic
  ultra-violet (UV) resonance lines, which have opacities that depend
  linearly on density, are directly affected by clumping only if
  individual clumps are \textit{optically thick}. However, recent
  results indicate that clumps are indeed thick in these lines
  \citep{Prinja10}, which then can lead to reduced line strengths and
  \textit{underestimates} of mass-loss rates if neglected in the
  analysis \citep[][]{Oskinova07, Sundqvist11}.}  The analysis here
examines the degree to which X-ray line profiles can provide a
mass-loss diagnostic that is relatively insensitive to clumping.

In single O stars without strong magnetic fields, X-rays are believed
to originate in embedded wind shocks associated with the LDI
\citep{Feldmeier97}, and the broad emission lines revealed by
high-resolution X-ray spectroscopy support this basic scenario
\citep{Kahn01, Cassinelli01, Cohen06}. These X-ray lines are often
observed as blue-shifted and asymmetric, characteristics stemming from
attenuation by bound-free absorption in the bulk wind
\citep{Macfarlane91}. As seen by an observer, X-ray photons emitted in
the receding part of the wind travel farther before escape, and thus
undergo more absorption, than those emitted in the advancing part.

For optically thin clumps, the amount of bound-free absorption is
proportional to the local density, and may thereby be used to put
additional constraints on mass-loss rates. Whilst initial analyses
\citep[e.g.,][]{Kramer03} required very low mass-loss rates to
reproduce the observed X-ray lines, more recent investigations with a
better account of the wind opacity and line blends \citep{Cohen10,
  Cohen10b, Cohen11} show that rates inferred from X-ray lines are
consistent with those derived from other diagnostics, \textit{if}
clumping is adequately accounted for in the other
diagnostics. However, a possible shortcoming of these X-ray analyses
is the assumption that clumps are optically thin, which if not met
would lead to an overestimate of the wind opacity, due to the
principal effect of \textit{porosity}.

Wind porosity models aiming to calculate X-ray line profiles have been
developed by, e.g., \citet{Feldmeier03}, \citet{Oskinova04}, and
\citet{Owocki06} (hereafter OC06). The first two of these studies
assumed the clumps to be \textit{radially oriented}, geometrically
thin shell fragments (`pancakes'), leading to a distinct anisotropic
form of the effective opacity. In OC06, on the other hand, the clumps
were assumed \textit{isotropic} to impinging radiation. Whereas
geometrically thin shell structures are indeed seen in one-dimensional
(1-D) LDI simulations, first attempts to construct 2-D LDI models
suggest that these shells break up via Rayleigh-Taylor or thin shell
instabilities into small clumps of similar angular and radial scales
\citep{Dessart03, Dessart05}. But these initial two-dimensional
simulations do not yet properly treat the lateral radiation transport
that might couple material, and so the degree of anisotropy of
instability generated structure in a fully consistent 3-D model is
still uncertain.

From the diagnostic side, OCO6 \citep[see also][]{Cohen08} argued that
for porosity to significantly affect the X-ray line formation,
unrealistically large porosity lengths, defined as the mean free path
between clumps (see Sect.~\ref{formalisms}), must be invoked. This
view is also supported by the above-noted recent attempts to derive
mass-loss rates directly from X-ray diagnostics. On the other hand,
\citet{Oskinova06} have argued that anisotropic clumps enhance
porosity effects, and lead to more-symmetric line profiles than if
assuming isotropic clumps, in general agreement with X-ray
observations.

This paper and its sequel (Leutenegger et al. 2011, hereafter
Paper\,II) further examine these issues. Building upon previous works
by \citet{Owocki01} and OC06, Sects.~\ref{opacities}
and~\ref{formalisms} develop a generalised formalism for synthesising
X-ray lines, including porosity as caused by either isotropic
(spherical, or randomly oriented) or anisotropic (flattened, radially
oriented) clumps. In addition, we generalise our models to treat an
ensemble of clumps of some distribution in optical depth, rather than
retaining the assumption that all clumps are \textit{locally}
identical.  Sect.~\ref{results} then systematically examines synthetic
X-ray line profiles and analyses porosity effects for isotropic and
anisotropic clumps, as well as for uniform and exponential clump
distributions. We discuss how the shape of the clumps affects the line
profiles in cases where porosity is important, and how this may be
used to put empirical constraints on the wind's clump geometry
(leaving detailed confrontation with observed spectra to Paper\,II).
Sect.~\ref{LDI} presents first X-ray line profiles calculated directly
from LDI simulations, using the 3-D patch method first developed by
\citet{Dessart02}.  Sect.~\ref{distribution} gives a physical
interpretation of the analytic porosity models presented, showing they
can be reconciled with a general statistical model derived for
stochastic transport in a two-component Markovian mixture of
immiscible fluids. Finally, in Sect.~\ref{discussion} we discuss our
results, compare them to other studies, and give our conclusions.
       
\section{Opacities in a clumped hot star wind}
\label{opacities}

\begin{figure}
\resizebox{\hsize}{!}{\includegraphics[angle=90]{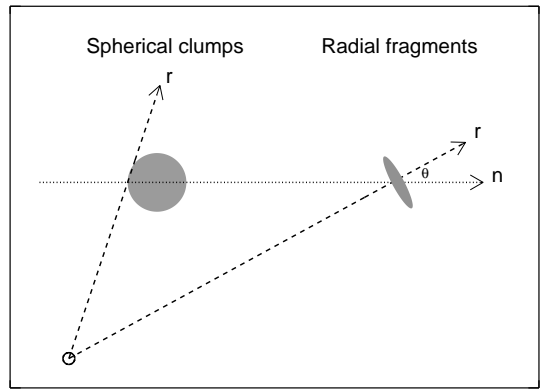}}
\caption{Schematic of photon of direction $\mathbf{n}$ impacting a
  spherical clump and a radially oriented ($\mathbf{r}/|r|$) shell
  fragment. The former gives an isotropic effective opacity whereas
  the latter gives an anisotropic effective opacity, since the
  projected surface area depends on direction cosine $\mu \equiv \cos
  \theta = \mathbf{n} \cdot \mathbf{r}/|r|$.}
\label{Fig:clumps}
\end{figure}

\begin{figure*}
\resizebox{\hsize}{!}{\includegraphics[]{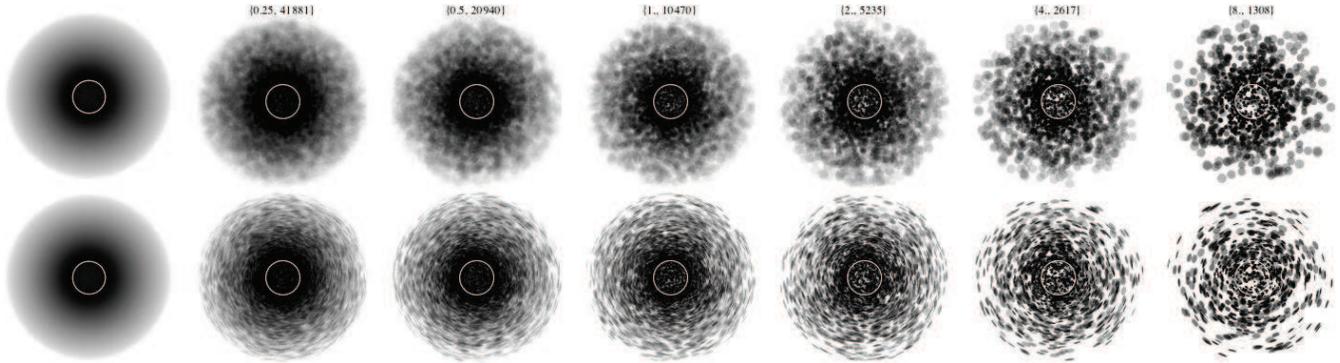}}
\caption{Back-lit rendition of randomly generated spherical (upper row)
  and radially compressed (lower row) clumps in a spherically
  expanding structured wind (columns 2-7), compared to a back-lit
  smooth wind (column 1) with $\tau_\star=1$ and the same total mass
  between an onset radius 1.05\,$R_\star$ and a maximum visualisation
  radius 5\,$R_\star$.  The terminal porosity length
  $h_\infty/R_\star$ increases from left to right, as given by the
  left header, but clump diameters all scale as $d_{\rm cl} =
  (0.2/\sqrt{\pi}) r$.  The right header gives the total number of
  clumps. The white circle represents the star, which also radiates
  with the same surface brightness as the background source.  Appendix
  A gives details on how these visualisations were generated.}
\label{Fig:rendition}
\end{figure*}

In our phenomenological model, we assume that the opacities in the
bulk wind can be described using a two-component medium consisting of
overdense `clumps' (denoted with cl) and a rarefied `inter-clump
medium'. The distribution of X-ray emitters in the shock-heated wind
is described in Sect.~\ref{Xray_emitters}. We neglect the inter-clump
medium's contribution to the opacities, an assumption well justified
for absorption of X-rays in O-star winds, due to the generally low
X-ray optical depths found for such stars \citep{Cohen10,
  Cohen11}. The volume filling fraction of the dense gas is $f_V$,
thus the local mean density is $\langle \rho \rangle = f_V \rho_{\rm
  cl}$, where the angle brackets denote spatial averaging.

\subsection{Optically thin clumps}

X-rays emitted in the wind are attenuated by bound-free absorption
depending \textit{linearly} on density. The local atomic mean volume
opacity per unit length is then $\langle \chi \rangle = \kappa \langle
\rho \rangle$, with mass absorption coefficient $\kappa$. By requiring
that the mean density of the clumped wind be equal to the density of a
corresponding smooth wind model, i.e. that $\langle \rho \rangle =
\rho_{\rm sm}$, one immediately recognises the well-known result that
for atomic processes depending linearly on density, the opacities are
not \textit{directly} affected by clumping as long as the individual
clumps remain optically thin. However, also linear-density opacities
can via a modified wind ionization balance be \textit{indirectly}
affected by optically thin clumping, but modelling such ionisation
equilibria is not a focus of the present paper.

As a comparison, for processes depending on the square of the density
(e.g. H$_\alpha$ in hot star winds), the opacities are always enhanced
compared to smooth models, by a factor given by the so-called clumping
factor $f_{\rm cl} \equiv \langle \rho^2 \rangle/\langle \rho
\rangle^2 = f_V^{-1}$, where the latter equality holds when the
inter-clump medium is neglected, whereby one obtains\footnote{In this
  context, we note that the porosity formalism presented in
  Sect.~\ref{formalisms}, although developed there for the specific
  case of X-ray line attenuation, is applicable also for continuum
  processes depending on $\langle \rho^2 \rangle$, for example thermal
  free-free emission, simply by exchanging the expression for $\langle
  \chi \rangle$.} $\langle \chi \rangle \propto \langle \rho^2 \rangle
= f_{\rm cl} \langle \rho \rangle^2 = f_V^{-1} \rho_{\rm sm}^2$.

\subsection{Porosity} 

It is important to realise that $\langle \chi \rangle$ may be
accurately used in radiative transfer models only in the limit of
optically thin clumps, $\tau_{\rm cl} \ll 1$. If this condition is not
satisfied for the investigated process, the radiative transfer becomes
more complex. For continuum processes such as the attenuation of
X-rays considered here, optically thick clumps lead to a local
self-shielding of opacity within the clumps, which in turn allows for
increased escape of radiation through porous channels in between the
clumps\footnote{For \textit{line} formation in a rapidly accelerating
  clumped medium, optically thick clumps lead to corresponding
  \textit{velocity} gaps, through which line photons may escape
  without ever interacting with the material \citep{Owocki08,
    Sundqvist10}. This is a consequence of the Doppler shift, leading
  to a picture wherein the clump length scales can be comparable to
  (or even larger than) the extent of the lines' resonance zones,
  which is a limit wherein the porosity formalism developed here is
  not directly applicable.}. The essential effect of such
\textit{porosity} is that the `effective' opacity of the medium
becomes lower than predicted by an optically thin clump model. Thus
porosity can mimic the symmetrising effects of reduced mass loss on
the X-ray line profiles. The purpose of this paper is to present a
formalism for quantifying this reduction and to show how clump
geometry and distribution affect X-ray line profile morphology in the
presence of porosity.

\section{A porosity formalism for effective opacity}
\label{formalisms}

In analogy with the atomic opacity, we may write the effective opacity
per unit length of a clump ensemble as \citep[e.g.,][]{Feldmeier03}
\begin{equation} 
\chi_{\rm eff} = n_{\rm cl}A_{\rm cl} P,
\label{Eq:chieff1}
\end{equation} 
where $n_{\rm cl}$ is the number density of clumps, $A_{\rm cl}$ is
the \textit{projected} area (the geometric cross-section) of a clump
for direction $\mathbf{n}$, and $P$ is the probability that a photon
impacting a clump gets absorbed. This probability obviously depends on
the clump optical depth, $P=P(\tau_{\rm cl})$, which we here
characterise by the average over all possible chord lengths $\ell$
through the clump for rays of direction $\mathbf{n}$,
\begin{equation}
  \tau_{\rm cl} = \frac{\int \tau_\ell \, dA_{\rm cl}}{A_{\rm cl}} = 
  \frac{\int \int \kappa \rho_{\rm cl} d \ell \, dA_{\rm cl}}{A_{\rm cl}} = 
  \frac{\kappa M_{\rm cl}}{A_{\rm cl}},
  \label{Eq:tauclform}
\end{equation}
where $M_{\rm cl}$ is the mass of the clump and the last equality
assumes that $\kappa$ is constant over the clump.

Another useful quantity is the local mean free path between clumps,
also known as the porosity length $h$ \citep{Owocki04},
\begin{equation}
  h \equiv \frac{1}{n_{\rm cl} A_{\rm cl}}.
  \label{Eq:hform}
\end{equation}
Using this definition, $\tau_{\rm cl}$ may be written
\begin{equation} 
  \tau_{\rm cl} = \frac{\kappa M_{\rm cl}}{A_{\rm cl}} = 
  \kappa (n_{\rm cl} M_{\rm cl}) h = \kappa \langle \rho \rangle h = 
  \langle \chi \rangle h,   
  \label{Eq:taucliso}
\end{equation}
whereby 
\begin{equation} 
\chi_{\rm eff} = \frac{P(\tau_{\rm cl})}{h} \ \ 
\Rightarrow \ \ 
\frac{\chi_{\rm eff}}{\langle \chi \rangle} = \frac{P(\tau_{\rm cl})}{\tau_{\rm cl}}. 
\label{Eq:chieffiso}
\end{equation}
For $P=1$, Eq.~\ref{Eq:chieffiso} returns the \textit{atomic opacity
  independent} result $\chi_{\rm eff}=1/h$, demonstrating how the
porosity length $h$ can also be interpreted as a photon's mean free
path in the limit of only optically thick clumps.  Further, for a
constant clump density Eq.~\ref{Eq:tauclform} yields $\tau_{\rm cl} =
\kappa \rho_{\rm cl} \ell_{\rm av}$, with average chord length
$\ell_{\rm av}$, which then recovers the commonly used form $h =
\ell_{\rm av}/f_{\rm V}$ for the porosity length.
 
\subsection{Isotropy vs. anisotropy} 

Note that the porosity length as defined in Eq.~\ref{Eq:hform} is a
strictly \textit{local} quantity. And because $\tau_{\rm cl} = \langle
\chi \rangle h$, this means that the (an)isotropy of the effective
opacity (Eq.~\ref{Eq:chieffiso}) depends only on $A_{\rm cl}$, and is
\textit{independent} of the spatial variation of $h$ associated with
the global wind expansion.  Thus, spherical clumps
(Fig.~\ref{Fig:clumps}), as well as randomly oriented clumps of
arbitrary shape, will have an isotropic effective opacity. For these
cases then, assuming spherical symmetry for the global wind expansion,
one may set $h = h(r)$ for all directions $\mathbf{n}$ impacting the
clump.
 
However, now let us consider a specific wind model in which the clumps
really are randomly distributed, but \textit{radially oriented},
geometrically thin shell fragments (`pancakes') \citep{Feldmeier03,
  Oskinova04}. In such a model, the projected clump area is $A_{\rm
  cl} \mathbf{n} \cdot \mathbf{r}/|r| = A_{\mu=1} |\mu|$, where we
here identify the projected area for a radially directed ray with the
area for an isotropic clump (Fig.~\ref{Fig:clumps}). 

This implies the effective opacity retains its basic form
(Eq.~\ref{Eq:chieffiso}) also for such anisotropic clumps, but the
clump optical depth becomes larger for oblique rays,
\begin{equation}
  \tau_{\rm cl}(r,\mu) = \langle \chi \rangle h(r,\mu) = \langle \chi \rangle h(r)/|\mu|, 
  \label{Eq:tauclaniso}
\end{equation}
where the isotropic case is recovered by setting $\mu=1$, and the
radial dependencies of the mean opacity and directional cosine have
been suppressed. This paper considers only isotropic and `pancake'
geometries, but the porosity formalism outlined above applies
generally to any clump geometry described by $A_{\rm cl} =
f(\mathbf{n})$, where $f$ is some function.

To illustrate such isotropic vs. anisotropic absorbing media,
Fig.~\ref{Fig:rendition} compares a random distribution of spherical
clumps and radially oriented pancakes, as illuminated by a uniform
background source (see Appendix A). Note that for visual clarity, we
extend these visualisations only to an outer radius 5\,$R_\ast$.  We
stress that Fig.~\ref{Fig:rendition} is for general illustration
purposes; the uniform background illumination is not a distribution of
X-ray emitters. Also, we repeat that though assumed in the
visualisations, spherical clumps are actually not a necessity for
obtaining isotropic porosity; the requirement for this is rather that
the clumps be randomly oriented, see above.

\subsection{Bridging laws for the effective opacity}
\label{bridge}

For attenuation with a given local clump optical depth $\tau_{\rm
  cl}=\tau_0$, let us assume the probability of absorption simply
takes the basic form $P=1-e^{-\tau_0}$. While formally exact only in
cases where all chord lengths across the clump are equal, this
expression is a suitable approximation that yields with
Eq.~\ref{Eq:chieffiso}
\begin{equation} 
\frac{ \chi_{\rm eff} }{\langle \chi \rangle } = 
\frac{1-e^{-\tau_0}}{\tau_0}. 
\label{Eq:bridgetauc}
\end{equation} 
This `single clump' bridging law now has the correct values in the limiting cases; it returns the atomic mean opacity when $\tau_0 \ll 1$ and is independent of it when $\tau_0 \gg 1$. And as discussed in the preceding section, the bridging law equation applies for both isotropic and anisotropic effective opacity models, however with different expressions for the clump optical depth. 

Assuming the effective mean free path scales as $\chi_{\rm
  eff}^{-1}=\langle \chi \rangle^{-1}+h$, an even simpler `inverse'
(or `Rosseland', see OC06) bridging law was invoked in OC06,
\begin{equation} 
\frac{\chi_{\rm eff}}{\langle \chi \rangle}= \frac{1}{1+\tau_{\rm 0}}, 
\label{Eq:bridgetauexp}
\end{equation} 
which also has the correct optically thin and thick limits.  By
considering the \textit{direction dependent} mean free path, this
inverse bridging law can be realised also for anisotropic models.

In OC06, the practical motivation for invoking
Eq.~\ref{Eq:bridgetauexp} was because the optical depth integral for
X-ray attenuation could then be solved analytically. However,
Sect.~\ref{distribution} shows it also happens to represent the
bridging law that follows from assuming the local clump optical depth
distribution function obeys Markovian statistics. Thus, the
single-clump and inverse bridging laws (Eq.~\ref{Eq:bridgetauc} and
Eq.~\ref{Eq:bridgetauexp}) differ in that the former assumes all
clumps have the same local optical depth (for a given direction),
whereas the latter averages over an exponential distribution in
$\tau_{\rm cl}$.

\subsection{Velocity stretch porosity}
\label{stretch}

OC06 assumed that the porosity length scales with the local radius,
but for mass-conserving clumps, such a scaling is only appropriate for
isotropic expansion. For clumps released into a radially expanding
stellar wind, the wind acceleration will `stretch' the clump spacing
in proportion to the wind velocity \citep{Feldmeier03}. The analysis
here assumes this velocity stretch form for both isotropic and
anisotropic porosity.

This distinction is most easily seen for the radially fragmented shell
model, in which the average radial separation between two shells is
$\Delta r = h(r)$.  For shells moving radially according to a
`$\beta$-velocity law', $w(r)= v(r)/v_\infty = (1-R_\star/r)^\beta$,
where $R_\star$ is the stellar radius, the separation is $h(r) =
h_\infty w(r)$, with the parameter $h_\infty$ representing the
asymptotic radial separation as $w \rightarrow 1$. For simplicity,
this paper assumes the prototypical value $\beta=1$. The quantity
$h_\infty/v_\infty$ represents the average time between two
consecutive shell passings at a fixed radial point in the wind
\citep{Sundqvist10}, which may also be interpreted as the inverse of a
`fragmentation frequency' $n_0$ \citep{Oskinova04, Oskinova06}.  For
example, $h_\infty=R_\star$ gives a fragmentation frequency $n_0 =
v_\infty/R_\star$ that is equal to the inverse of the wind flow time.

\subsection{X-ray line transfer in porosity models} 
\label{Xray_emitters}

To compute X-ray emission line profiles, we solve the standard formal
integral of radiative transfer, using a customary $(p,z)$ coordinate
system and following the basic procedure described in \citet{Owocki01}
for the distribution of X-ray emitters. Since our primary interest
here is the \textit{shapes} of the lines, all resulting flux profiles
have been normalised to a unit maximum, ${\cal F}_{\rm x}^{\rm norm} =
{\cal F}_{\rm x}/\rm Max({\cal F}_{\rm x})$. For simplicity we assume
that the X-ray emission begins at a certain onset radius $R_0=1.5 \,
R_\star$, and is constant beyond it. This onset radius is consistent
with that typically predicted by conservative, self-excited LDI
simulations \citep{Runacres02}, and both $R_0 \approx 1.5 R_\star$ and
a constant X-ray filling factor are supported by observations
\citep{Leutenegger06, Cohen06, Cohen11}. Moreover, both LDI
simulations and the observed X-ray luminosities indicate that only a
very small mass fraction, less than 1\,$\%$, of the stellar wind is
shock-heated to X-ray emitting temperatures at any given
time. Simulations and observed lack of X-ray variability further
indicate that there are numerous sites of X-ray emission distributed
throughout the wind, justifying our assumption of a smoothly
distributed X-ray emitting plasma above the onset radius.  Note
though, that $R_0$ does not necessarily equal the clump onset radius
$R_{\rm cl}$, which observations typically indicate is located much
closer to the photosphere \citep[][see also Fig.~\ref{Fig:rendition}
  for a visualisation]{Puls06, Cohen11}. But in the velocity stretch
porosity models, we have verified that the exact value of this $R_{\rm
  cl}$ is not important for the resulting X-ray line profiles.

The absorption of X-rays emitted at position $Z_{\rm e}$ along a ray
with impact parameter $p$ is given by the optical depth integral
\begin{equation}
  \tau(p,z_{\rm e}) = \int_{z_{\rm e}}^\infty \chi_{\rm eff}(z,p) dz,  
  \label{Eq:tauinte}
\end{equation} 
where the \textit{effective} opacity accounts for any porosity.  The
opacity in a smooth or optically thin clump model, due purely to the
atomic mean opacity $\langle \chi \rangle$, is proportional to the
mass-loss rate $\dot{M}$ of the star and here characterised by a
fiducial optical depth $\tau_\star = \dot{M} \kappa/(4 \pi R_\star
v_\infty)$, with wind terminal speed $v_\infty$. To evaluate
Eq.~\ref{Eq:tauinte} for our porosity models, the only additional
input parameter required is the porosity length $h$ (see
Eqs.~\ref{Eq:taucliso}-\ref{Eq:bridgetauexp}). This holds for
isotropic as well as anisotropic effective opacity models, and for the
single-clump as well as the inverse bridging laws.\footnote{Source
  codes to all X-ray line porosity models presented in this section
  are publicly available at (package {\sc windprof}) {\sffamily
    http://heasarc.gsfc.nasa.gov/docs/xanadu/xspec/models/wind
    prof.html}, where models for the broadband absorption of X-rays
  \citep[package {\sc windtabs},][]{Leutenegger10} also can be found.}

\section{X-ray line profiles from analytic porosity models}
\label{results}

\begin{figure*}
\resizebox{\hsize}{!}{\includegraphics[angle=90]{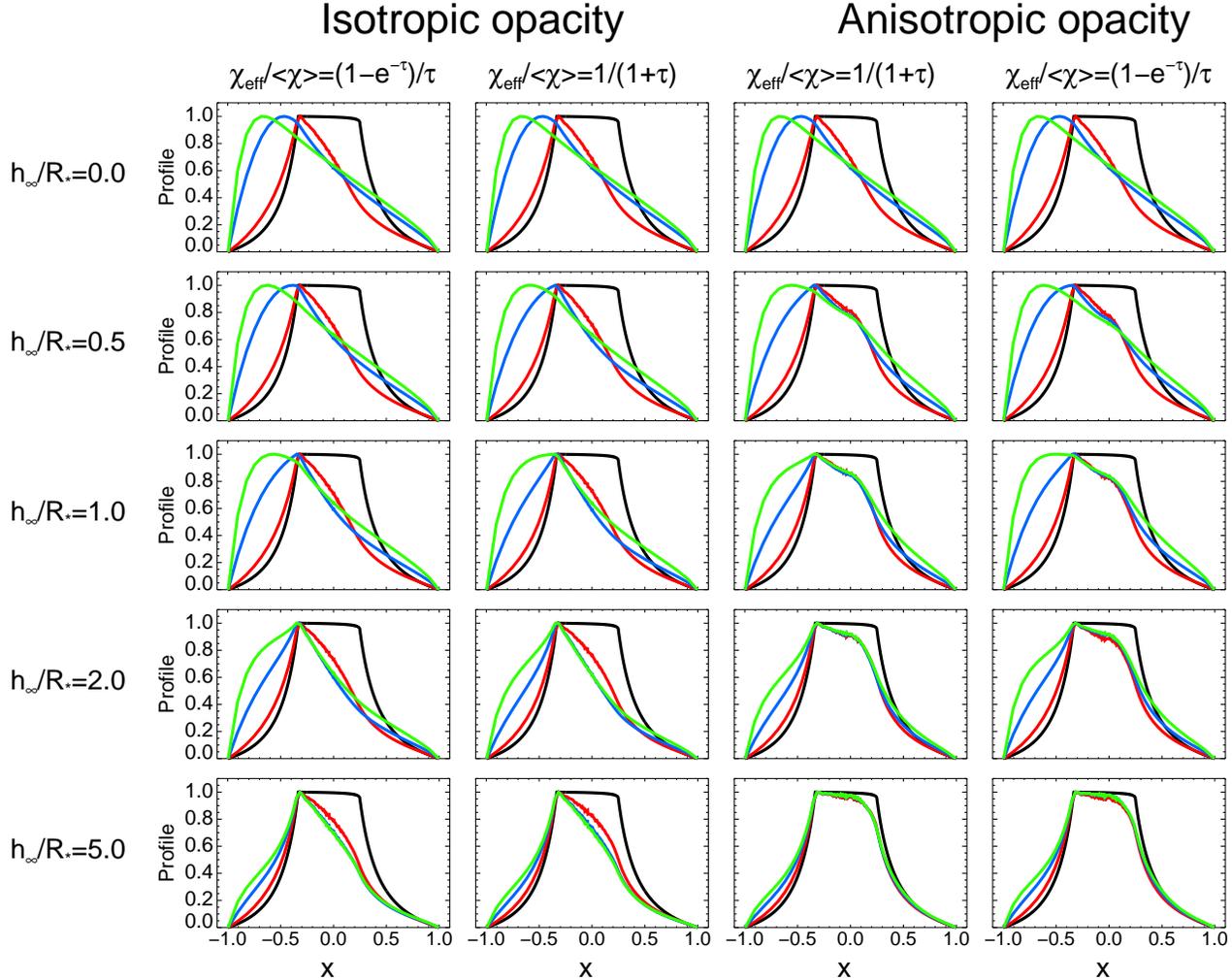}}
\caption{Synthetic X-ray line profiles for the `single clump',
  $\chi_{\rm eff}/\langle \chi \rangle = (1-e^{-\tau})/\tau$, and the
  `inverse', $\chi_{\rm eff}/ \langle \chi \rangle = 1/(1+\tau)$,
  effective opacity bridging laws and for different porosity length
  parameters $h_\infty/R_\star$, as labelled. All models assume an
  onset radius for the X-ray emission $R_0 = 1.5 R_\star$. The
  abscissae display the dimensionless wavelength
  $x=(\lambda/\lambda_0-1)c/v_\infty$, with $\lambda_0$ the
  line-centre wavelength, and the ordinates display the normalised
  flux. Black, red, blue, and green dashed lines have an optical depth
  parameter $\tau_\star=0.01, 1, 5, 10$, respectively; for a
  non-colour separation, an increased $\tau_\star$ means a more
  blue-shifted peak flux. Note that we have set $h_\infty=0$ (the
  uppermost panel) to be equivalent to assuming only optically thin
  clumps.}
\label{Fig:profs}
\end{figure*}

Fig.~\ref{Fig:profs} displays synthetic X-ray line profiles calculated
using the four possible combinations of isotropic vs. anisotropic
effective opacity and single-clump vs. inverse bridging laws. The
figure clearly shows that profiles calculated using the two different
bridging laws are very similar (see also Fig.~\ref{Fig:tau5}, as well
as Fig.~1 in OC06), despite representing two very different clump
optical depth distributions (Sect.~\ref{bridge}). This indicates that
the effects of porosity on X-ray line profiles are not very sensitive
to the specific local distribution in $\tau_{\rm cl}$.  We discuss
this important result further in Sect.~\ref{distribution}.

\begin{figure}
\resizebox{\hsize}{!}{\includegraphics[angle=90]{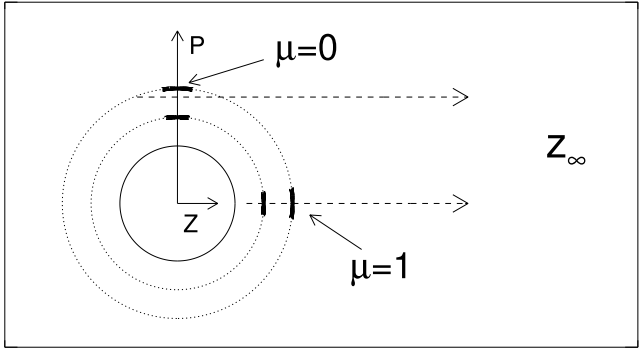}}
\caption{Illustration of the `venetian blind' effect seen in porosity
  models using an anisotropic effective opacity. The dashed arrowed
  lines represent two different $p$-rays and the observer is assumed
  to be located at $z_\infty$.}
\label{Fig:venetian}
\end{figure}

\begin{figure}
\resizebox{\hsize}{!}{\includegraphics[angle=90]{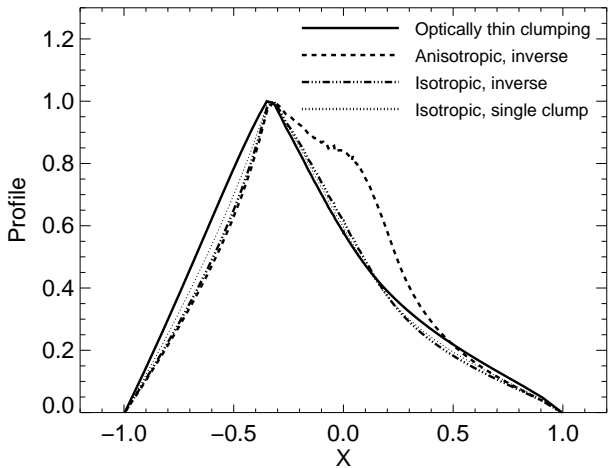}}
\caption{Line profiles for $h_\infty=R_\star$ and
  $\tau_\star=2.5$, using different effective opacity laws, as
  labelled.}
\label{Fig:tau5}
\end{figure}

The second key feature of Figs.~\ref{Fig:profs} and~\ref{Fig:tau5} is
the prominent `bump' visible close to line centre in profiles
calculated with anisotropic effective opacity. Conceptually, we may
understand this as a `venetian blind' effect (Fig.~\ref{Fig:venetian},
see also \citeauthor{Feldmeier03} 2003); since the fragmented shells
are radially oriented, the blinds are closed for radial photons, but
open up for more tangential ones. This leads to increased escape for
photons emitted close to line centre, since the line emission
wavelength scales with direction cosine $\mu$ as $x=-\mu w$
\citep[e.g.,][]{Owocki01}.

Another way to look at this effect is to consider the optical depth
integral for anisotropic effective opacity in the $\tau_{\rm cl} \gg
1$ limit,
\begin{equation} 
  \tau(p,z) = \int_{z_{\rm e}}^\infty \chi_{\rm eff}(z,p) dz
  \approx \int_{z_{\rm e}}^\infty \frac{|\mu|}{h(r)} dz,  
  \label{Eq:counting}
\end{equation} 
which shows that, since $dr = \mu dz$, the optical depth in this limit
is set simply by counting up the number of porosity lengths. In the
plane-parallel limit of radially oriented, geometrically thin but
optically thick fragments, \textit{all} tangential ($\mu=0$) photons
would escape. However, due to sphericity effects (i.e., that $\mu$
increases as the photon propagates through the wind) also photons
emitted initially in the tangential direction will suffer some
absorption (see Fig.~\ref{Fig:venetian}).  Thus the end result is not
complete transmission, but a characteristic bump stemming from the
reduced integrated optical depth for photons emitted around $x \approx
0$. For isotropic porosity, on the other hand, no $\mu$ factor enters
in Eq.~\ref{Eq:counting}, and therefore no bump appears in these
profiles.

This quite distinct and \textit{systematic} difference in the shape
around line centre between models with isotropic and anisotropic
effective opacity is a key result of the present analysis. Indeed, one
can us this difference to set empirical constraints on the clump
geometry by confronting synthetic X-ray spectra with observed ones, as
will be done in Paper\,II.

Generally, Fig.~\ref{Fig:profs} confirms earlier results by OC06 that
in order to achieve a significant effect on the profiles, rather large
porosity lengths, $h_\infty > R_\star$, are required. However,
Fig.~\ref{Fig:tau5} reveals that for a value of $\tau_\star=2.5$,
representative of the prototypical O supergiant $\zeta$ Pup
\citep{Cohen10}, the anisotropic porosity model displays significantly
higher flux around line centre than the other profiles, also for
$h_\infty=R_\star$. Further comparisons show that, for the parameters
used in Fig.~\ref{Fig:tau5}, the isotropic porosity model is well
matched by a corresponding optically thin clump model with
$\tau_\star$ reduced by $\sim$\,20 percent (a detailed
quantification of this trade-off will be provided in Paper\,II). In
contrast, such a simple optical depth reduction does not reproduce the
line-centre region in anisotropic porosity models, i.e.  there is no
simple trade-off, or degeneracy, between mass-loss rate and
anisotropic porosity. Thus, whereas it will be difficult to
distinguish between optically thin clumps and moderate isotropic
porosity with somewhat higher $\tau_\star$'s, careful line-fitting to
observations should be able to identify, or refute, an anisotropic
porosity.

Finally, for very large porosity lengths, giving $\tau_{\rm cl} \gg 1$
in a large portion of the wind, the profile shapes do indeed become
quite independent of atomic opacity. In the bottom panel of
Fig.~\ref{Fig:profs}, profiles computed using an anisotropic effective
opacity are actually very similar to the flat-topped profiles stemming
from computations without any X-ray absorption, whereas profiles
computed using an isotropic effective opacity retain a certain degree
of asymmetry. This result differs from that found in OC06, wherein
very near symmetry was achieved for isotropic profiles with very large
porosity lengths, and comes about due to the $h \propto v$ scaling
adopted here, which implies significantly shorter porosity lengths in
the lower wind regions than the $h \propto r$ used in OC06. However,
for more moderate values of $h_\infty$ (top three panels in
Fig.~\ref{Fig:profs}), all profiles are opacity dependent, as expected
since in the formation of these lines, porosity is a secondary effect.

\begin{figure}
\resizebox{\hsize}{!}{\includegraphics[angle=90]{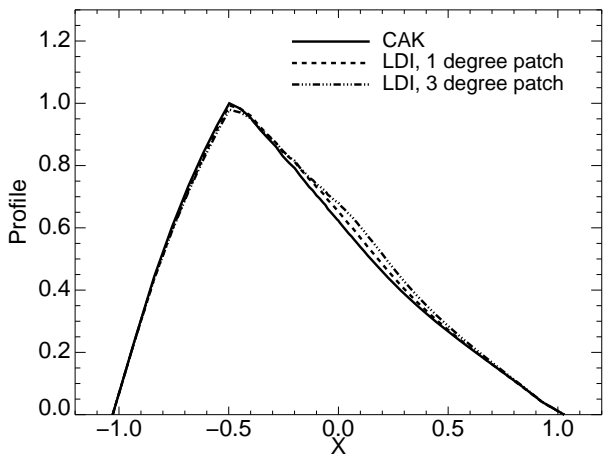}}
\caption{Line profiles for $\tau_\star=2.5$, calculated from a smooth CAK model 
         and structured LDI models with patch sizes 1 and 3 degrees (see text), 
         as labelled.}
\label{Fig:tau2.5_LDI}
\end{figure}

\section{X-ray line profiles from LDI simulations}
\label{LDI}

The above simplified analytic models demonstrate that the effect of
optically thick clumps on wind continuum absorption can be well
characterised in terms of the wind porosity length, parametrised here
by the asymptotic value $h_\infty$ and assuming a radial variation set
by the velocity-stretch form, $h(r) = h_\infty (1-R_\star/r)$
(Sect.~\ref{stretch}).

But inspection of LDI simulations suggests a substantially steeper
radial variation for the separation between instability-generated wind
clumps. Even in 1-D models in which the separation can become of the
order of a stellar radius in the outer wind (implying $h_\infty
\approx R_\ast$), the initial clump structure formed near the onset
radius $r \approx 1.5 R_\ast$ tends to have a much smaller separation
scale, of the order of the Sobolev length $L_{\rm sob} \approx (v_{\rm
  th}/v) R_\ast \approx 0.01 R_\ast$, where $v_{\rm th}$ is the ion
thermal velocity. The sharp increase in separation from this onset
comes not just from velocity stretching from the overall wind
acceleration, but also from {\em collisional merging} of clumps with
substantial radial velocity dispersion. For a given asymptotic
porosity length $h_\infty$, LDI models thus tend to have smaller
inner-wind porosity lengths than assumed in the simple velocity
stretch scaling. Since it is in this inner wind region that clumps can
become optically thick for X-rays, this suggests that LDI models will
show even weaker porosity effects than implied by the analytic
profiles shown in Figs.~\ref{Fig:profs} and~\ref{Fig:tau5}.  To
demonstrate this explicitly, we now present some first sample
calculations of X-ray profiles computed from 1-D LDI simulations that
are phased randomly among 3-D patches of a parametrised angular size
\citep[as first developed in][]{Dessart02}. The details of this patch
geometry as implemented in our radiative transfer code are given in
\citet{Sundqvist11}. Here we adapt this code to synthesise X-ray line
profiles by making the following assumptions: The X-ray emission is
assumed to have a fixed spatial form independent of wind structure,
scaled in proportion to the density squared of a smooth CAK wind and
with an onset radius $R_0 = 1.5 R_\ast$. The bound-free absorption is
then calculated directly from the structured LDI simulation presented
in \citet{Sundqvist11} (computed following \citeauthor{Feldmeier97}
1997).

Fig.~\ref{Fig:tau2.5_LDI} plots X-ray line profiles for the same wind
optical depth used for analytic models in Fig.~\ref{Fig:tau5}, namely
$\tau_\ast = 2.5$. The curves compare a smooth CAK model to structured
LDI models with patch sizes of 1 and 3 degrees. The overall shapes
agree well with corresponding non-porous analytic models, except for
small differences due to the fact that the CAK velocity law does not
exactly follow the phenomenological $\beta=1$ law. But the key point
is that the LDI profiles are very similar to the CAK profile, implying
little porosity effect; however, we note that for a 3-degree patch
size, there is a small, but noticeable, `bump' around line centre,
presumably associated with the anisotropic (pancake) nature of the 
clumped structure in such models.

These results confirm that porosity is as marginal in LDI simulations
as in corresponding analytic stretch-porosity models with comparable
asymptotic clump separations (i.e., $h_\infty \approx
R_\star$). Moreover, in 2-D LDI models \citep{Dessart03, Dessart05},
clumps can also be broken up by shearing and associated effects,
leading generally to more, smaller, and less optically thick clumps,
characterised by even smaller porosity lengths. Future work will
implement our radiative transfer tools also in such genuinely multi-D
instability models.

\section{A physical interpretation of the effective opacity bridging laws} 
\label{distribution}

As already noted, a basic difference between the two bridging laws
adopted in this paper is that one (Eq.~\ref{Eq:bridgetauc}, the
`single clump' law) assumes a locally constant clump optical depth for
all clumps, whereas the other (Eq.~\ref{Eq:bridgetauexp}, the
`inverse' law) represents a certain distribution in $\tau_{\rm
  cl}$. Before discussing porosity effects on the synthetic X-ray line
profiles in Sect.~\ref{discussion}, this section examines the nature
of this distribution.
 
\subsection{Distribution laws for $\tau_{\rm cl}$}

To investigate how a distribution of clump optical depths affects
effective opacity scalings, let us assume that the ratio of the
effective to the mean opacity scales as the ratio of an effective
clump optical thickness to its mean,
\begin{equation}  
    \frac{\chi_{\rm eff}}{\langle \chi \rangle} = 
    \frac{\tau_{\rm eff}}{\langle \tau \rangle} = 
    \frac{ \int_0^\infty \tau f(\tau) \frac{1-e^{-\tau}}{\tau}d\tau }
    { \int_0^\infty \tau f(\tau)d\tau },
    \label{Eq:chieffdist}
\end{equation}
where we for clarity have dropped the indices on the clump optical
depths.  Eq.~\ref{Eq:chieffdist} introduces $f(\tau)$, the normalised
distribution function of clumps, and $\tau_{\rm eff}$, the
distribution weighted mean of the clump optical depth. Selecting a
weighing function $(1-e^{-\tau})/\tau$ ensures that the single clump
bridging law (Eq.~\ref{Eq:bridgetauc}) is recovered from
Eq.~\ref{Eq:chieffdist} when the distribution function is a Dirac
delta function, $\delta(\tau-\tau_0)$.

Let us now choose a specific distribution function of the exponential
form
\begin{equation}  
  f(\tau) = \frac{e^{-\tau/\tau_0}}{\tau_0}, 
  \label{Eq:distexp}
\end{equation}
which has a mean value $\tau_0$.  Using Eq.~\ref{Eq:distexp} in
Eq.~\ref{Eq:chieffdist} yields directly the inverse bridging law
Eq.~\ref{Eq:bridgetauexp}, but with $\tau_0$ now only representing the
\textit{mean} clump optical depth, rather than a unique one as in the
exponential bridging law Eq.~\ref{Eq:bridgetauc}. Again, despite the
large difference between this clump distribution and the one assuming
a constant $\tau$, the two bridging laws give similar results, as
demonstrated in Sect.~\ref{results}.

\subsection{Connecting Markovian statistics to exponentially distributed clumps} 
We now show that the bridging law resulting from this exponential
distribution turns out to be a special case of a general scaling
derived for a stochastic mixture of two fluids that follow Markovian
statistics \citep[e.g.,][]{Levermore88, Pomraning91}.  The Markov
assumption is that the future state of the system only depends on its
present state, and not on its history. Along any given ray through the
medium, if the fluid is of component $0$ at location $s$, the
probability of it being of component $1$ at $s + ds$ is $P_{0,1}ds$,
where $P_{0,1}$ is independent of how far back along the ray the last
transition (from fluid $1$ to $0$) occurred. Under this assumption,
the length scales traveled within the fluid components are random
variables described by Poisson distributions, with $P_{0,1}$
identified as the inverse of $\ell_0$, the \textit{mean} distance a
photon travels along the ray in fluid $0$ before finding itself in
component $1$. A similar definition applies for $\ell_1$.

For such a two-component Markov model with opacities $\chi_0$ and
$\chi_1$, \citet{Levermore88} derived for the effective opacity in the
pure absorption case
\begin{equation}
  \chi_{\rm eff} = \frac{\langle \chi \rangle +
    \chi_0\chi_1 \ell_{\rm c}}{1+(p_0\chi_1+p_1\chi_0) \ell_{\rm c}},
  \label{Eq:chiefflever0}
\end{equation}
where $p_i \equiv \ell_i/(\ell_0+\ell_1)$, $\langle \chi \rangle =
p_0\chi_0 + \chi_1p_1$ is the mean opacity, and $\ell_{\rm c} \equiv
\ell_0 \ell_1/(\ell_0+\ell_1)$ is the correlation length
\citep{Pomraning91}.  Identifying the clumps in our model with
component $1$, and assuming the inter-clump medium to be void
($\chi_0=0$), we find for this `clump+void Markov model'
\begin{equation}
  \frac{\chi_{\rm eff}}{\langle \chi \rangle} = 
  \frac{1}{1+p_0 \chi_1 \ell_{\rm c}} = 
  \frac{1}{1+\langle \chi \rangle p_0 \ell_0} = 
  \frac{1}{1+\langle \chi \rangle h}.  
  \label{Eq:chiefflever}
\end{equation}
Here $p_0$ represents the probability that a photon is in the void
medium, while $\ell_0$ is the distance the photon travels in the void
before encountering a clump; the product $p_0 \ell_0$ thus represents
the photon mean free path in the case of optically thick clumps, which
is also the porosity length $h$, as given by the final equality in
Eq.~\ref{Eq:chiefflever}. Comparison with Eq.~\ref{Eq:bridgetauexp}
then shows that the effective opacity bridging law for an exponential
clump optical depth distribution is equivalent to that for this
statistical clump+void Markov model.

Indeed, recalling that the Markov transport is defined along a given
ray, we may make the same identification for the anisotropic porosity
model, with the photon mean free path along the ray then being scaled
by $1/|\mu|$.

\subsection{Exponentially truncated power-law distributions}

While there are not many observational constraints on the distribution
of clumps in a hot star wind \citep[see however][]{Lepine99,
  Dessart05b}, the above identification with the Markov model at least
places our porosity models on a robust and well-known statistical
ground.  The inverse bridging law should therefore be an appropriate
standard choice for porosity applications such as the X-ray line
formation considered here, but perhaps also for, e.g., porosity
moderated continuum driven wind models of stars formally exceeding the
Eddington luminosity, as investigated by \citet{Owocki04}. Indeed,
although not explicitly studied in that paper, we note that the Markov
model represents a special case of the exponentially truncated
power-law distribution of clumps considered in \citet{Owocki04},
namely the one with power index $\alpha_p=2$. Thus, reasonable
extensions of the two canonical distributions studied here could
readily be done by using some other power index variant given in
\citet{Owocki04}.

\section{Discussion and conclusions}
\label{discussion}
 
\subsection{Isotropy or anisotropy} 

Let us next compare our analysis to that by \citet{Oskinova06}. These
authors also pointed out the differences between isotropic and
anisotropic effective opacity and carried out a comparison, however
only for the specific case of $\tau_\star=10$ and a fragmentation
frequency (see Sect.~\ref{stretch}) $n_0 =1.4 \times 10^ {-4}\ \rm
s^{-1}$. Taking the parameters for the O supergiant $\zeta$ Pup
adopted in \citeauthor{Oskinova06}, this corresponds to
$h_\infty=v_\infty/n_0=4.24 R_\star$, which in turn roughly
corresponds to the bottom panel in our Fig.~\ref{Fig:profs}. Indeed,
the profiles displayed in that panel agree well with those in Fig.~16
of \citeauthor{Oskinova06}; both figures illustrate that for such very
large porosity lengths, profiles computed using anisotropic effective
opacity are nearly symmetric (Sect.~\ref{results}).

This comparison suggests an overall good agreement among the results
found by the different groups.  But as shown in Sect.~\ref{results},
anisotropic porosity line profiles, with their characteristic `bump'
at line centre, are qualitatively different than isotropic porosity or
optically thin clumping profiles. Thus the good statistical fits
presented for $\zeta$ Pup by \citet{Cohen10}, without invoking
porosity, seem somewhat contradictory to the good visual fits
presented by \citet{Oskinova06}, using models with moderate
anisotropic porosity $h_\infty \approx R_\star$. Paper\,II will
further examine and quantify these differences between anisotropic
porosity on the one hand, and isotropic porosity or optically thin
clumping on the other.

\subsection{Is porosity important for X-ray line mass-loss diagnostics?}
 
The LDI simulations presented in Sect.~\ref{LDI} indicate small
porosity lengths and negligible porosity effects on X-ray line
profiles. Such small porosity lengths also have some indirect
empirical support. Namely, the mass-loss rate derived for $\zeta$ Pup
by \citet{Cohen10}, without invoking porosity, is only marginally
lower than the \textit{upper limit} mass-loss rate derived by
\citet{Puls06}, by assuming an unclumped outermost radio emitting
wind, while allowing for clumping in the intermediate and lower
wind. Because of the trade-off between porosity and mass-loss rate
then, if porosity lengths large enough to significantly affect the
X-ray line profiles were to be adopted, $h_\infty > R_\star$, the
inferred X-ray mass-loss rate would be \textit{higher} than this upper
limit. That is, such multiwavelength considerations indicate that a
significant porosity effect on X-ray based mass-loss rates is
incompatible with diagnostic results from other wavebands.

Overall, we thus conclude that porosity effects on X-ray line profiles
are likely to be, at most, a marginal effect in typical O stars. This
is supported also by the low optical depths found for $\zeta$ Pup as
well as for the even denser wind of HD93129A \citep{Cohen10,
  Cohen11}. Since most O stars will have characteristic $\tau_\star$'s
significantly lower than these, porosity effects should be
negligible. The upshot is that X-ray line analysis may indeed provide
the best available `clumping insensitive' diagnostic of O star
mass-loss rates.

\section*{Acknowledgments}

This work was supported in part by NASA ATP grant NNX11AC40G.
D.H.C. acknowledges support from NASA ADAP grant NNX11AD26G to
Swarthmore College. We thank A. Feldmeier for providing the
instability simulations discussed in Sect.~\ref{LDI}, and for
suggesting the exploration of Markov models discussed in
Sect.~\ref{distribution}.

\appendix 

\section{Generation method for clumped medium illustrations}

The clumped medium illustrations in Fig.~\ref{Fig:rendition} were
generated by following the radial expansion of mass-conserving
clumps. As with the analogous illustration in Fig.~3 of OC06, we
assume the clump scale $l$ increases in proportion to the local
radius,
\beq
l(r) = l_{\ast} \frac{ r}{R_{\ast}} \, .
\label{ldef}
\eeq However, instead of the OC06 assumption of a purely isotropic
(`Hubble-law') velocity expansion $v \sim r$, we now use a standard
$\beta=1$ wind velocity law. For clumps of projected area $A_{\rm cl}
= l^{2} \propto r^{2}$ and local volume density $n_{cl} \propto 1/(v
r^{2})$, this gives the associated radial variation of the porosity
the desired `velocity-stretch' form,
\beq
h(r) = \frac{1}{n_{\rm cl} A_{\rm cl}} = h_{\infty} \frac{v(r)}{v_{\infty}} 
=  h_{\infty} (1-R_{\ast}/r) \, .
\eeq
For specified clump parameters $l_{\ast}$ and $h_{\infty}$, the clump
number density is thus given by
\beq
n_{cl} (r) = \frac{1}{h A_{\rm cl}} = 
\frac{v_{\infty} R_{\ast}^{2}}{h_{\infty} l_{\ast}^{2}} \, \frac{1}{r^{2} v(r)}
\, .
\eeq
Note that, unlike the OC06 isotropic expansion model, the clump volume
filling factor in this velocity-stretch scaling is {\em not constant},
but varies spatially as $f_{\rm V} \propto n_{\rm cl} l^{3} \propto
r/v(r)$, which actually is quite consistent with derived observational
constraints \citep[e.g.,][]{Puls06}.

The cumulative number of clumps up to a radius $r$ above the clump
onset radius $R_{\rm cl}$ is
\beqa
N(r) &=& 4 \pi \int_{R_{\rm cl}}^{r} n_{\rm cl} r'^{2} \, dr'
\nonumber \\ &=& \frac{4 \pi R_{\ast}^{3}} {h_{\infty} l_{\ast}^{2}} \, 
\left [ \frac{r-R_{\rm cl}}{R_{\ast}}  
 + \ln \left ( \frac{r-R_{\ast}}{R_{\rm cl} -R_{\ast}} \right ) 
\right ].
\eeqa
For a specified outer radius $R_{\rm max}$, the total number of clumps
is $N_{\rm tot} = N(R_{\rm max})$.  Since Fig.~\ref{Fig:rendition}
uses fixed parameters $l_{\ast}/R_{\ast} = 0.1$, $R_{\rm
  cl}/R_{\ast}=1.05$, and $R_{\rm max}/R_{\ast} = 5$, this number
scales with $1/h_{\infty}$, and ranges from $N_{\rm tot}=1\,308$ for
the largest porosity length $h_{\infty}/R_{\ast} = 8$ in the rightmost
column, to $N_{tot} = 41\,881$ for the least porous case
$h_{\infty}/R_{\ast} = 0.25$ in column 2.

A random set of $N_{\rm tot}$ clumps with the required statistical
distribution in radius can now be drawn simply by generating a set of
pseudo-random numbers ${\cal R}_{\rm i}$ over the range $[0,1]$, and
inverting the normalised cumulative distribution function $F(r) \equiv
N(r)/N_{tot} = \cal R_{\rm i}$ to find the radius $r$,
\beq
r({\cal R}_{\rm i}) = R_{\ast} \left ( 1 + {\rm ProductLog} [\exp(C_{\rm cl} + C_{\rm max} {\cal R}_{\rm i})] 
\right )
\, ,
\label{roff}
\eeq
with the constants given by
\beq
C_{\rm cl} ~~~ = R_{\rm cl}/R_{\ast} -1 + \ln( R_{\rm cl}/R_{\ast} -1) 
\eeq
\beq
C_{\rm max} = (R_{\rm max}-R_{\rm cl})/R_{\ast} 
+ \ln \left ( \frac{R_{\rm max}-R_{\ast}}{R_{\rm cl} - R_{\ast}} \right ) \,
.
\eeq
Likewise, we use the assumed statistical spherical symmetry and 
generate the clump angle coordinates in azimuth $\phi$ and 
colatitudinal cosine $\mu$ through additional pseudo-random
numbers,
\beq
\phi_{i} = 2 \pi {\cal R}_{i+ N_{tot}} ~~ ; ~~ \mu_{i} = 2 {\cal
R}_{i+2 N_{\rm tot}} - 1 \, .
\eeq
In this system it is most convenient to assume the clumps are viewed
from above the $\mu=1$ pole, with thus $\mu>0$ ($\mu < 0$)
representing clumps in the foreground (background) hemisphere.

With the random set of clump positions in hand, the clump sizes are
set by Eq.~(\ref{ldef}), with all cases in Fig.~\ref{Fig:rendition}
using $l_{\ast}/ R_{\ast} = 0.1$. To give the associated projected
area $l^{2}= A_{\rm cl} = \pi d_{\rm cl}^{2}/4$, the clump diameters
are set to $d_{\rm cl} = 2 l/\sqrt{\pi}$.

For the spherical clumps in the upper row of Fig.~\ref{Fig:rendition},
the transparency of each individual projected clump disk area is set
by $\exp(-\tau_{cl})$, where the surface-averaged clump optical depth
is $\tau_{cl} = \tau_{\ast} h_{\infty} R_{\ast}/r^{2}$.  For the
radially compressed clumps in the lower row, the associated clump
optical depths are increased by $1/|\mu|$; their projected areas are
reduced through foreshortening their radial extent by a factor
$|\mu|$, while keeping their radially perpendicular extent equal to
the local clump diameter $d_{\rm cl}$.

Finally, clumps in the back hemisphere ($\mu < 0$) that are behind the
star (with $r \sqrt{1-\mu^{2}} < R_{\ast}$) are simply not drawn.
This effectively means the clumps directly in front of the star appear
as if illuminated by a stellar surface brightness equal to the back
illumination of the clumps outside the stellar limb.


\bibliography{sundqvist}

\end{document}